%31-mar-98  
%%%%%%%%%%%%%%%%%%%%%%%%%%%%%%%%%%%%%%%%%%%%%%%%%%%%%%%
%%%%%%%%%%%%%%%%%%%%%%%%%%%%%%%%%%%%%%%%%%%%%%%%%%%%%%%
%%                                                   %%
%%                (Plain) TeX file of                %%
%%                                                   %%
%%    A Model--Independent and Rephase--Invariant    %%
%%         Parametrization of CP--Violation          %%
%%                                                   %%
%%                         by                        %%
%%                                                   %%
%%          D. COCOLICCHIO and M. VIGGIANO           %%
%%                                                   %%
%%    Preprint Sezione INFN Milano IFUM 609-FT/97    %%
%%                                                   %%
%%%%%%%%%%%%%%%%%%%%%%%%%%%%%%%%%%%%%%%%%%%%%%%%%%%%%%%
%%%%%%%%%%%%%%%%%%%%%%%%%%%%%%%%%%%%%%%%%%%%%%%%%%%%%%%
\magnification=1200
\def\nonumfirst{\nopagenumbers
                \footline={\ifnum\count0=1\hfill
                           \else\hfill\folio\hfill
                           \fi}}
\nonumfirst
%\pageno=1
%\nopagenumbers
\magnification=1200
%\input macro.tex
%%
%part of decioMACRO
\def\singlespace{\baselineskip 12 pt}

\def\oneandahalfspace{\baselineskip 18pt}
\def\blankline{\vskip 12 pt\noindent}
\def\blankblankline{\vskip 18 pt\noindent}
\def\secto#1\endsecto{\vskip 20pt {\bf #1}\vskip 7pt\nobreak}
\global\newcount\refno \global\refno=1
\newwrite\rfile

\def\ref#1#2{\hbox{[\hskip 2pt\the\refno\hskip 2pt]}\nref#1{#2}}
\def\nref#1#2{\xdef#1{\hbox{[\hskip 2pt\the\refno\hskip 2pt]}}%
\ifnum\refno=1\immediate\openout\rfile=refs.tmp\fi%
\immediate\write\rfile{\noexpand\item{\noexpand#1\ }#2.}%
\global\advance\refno by1}

\def\semi{;\hfil\noexpand\break}
\def\demi{:\hfil\noexpand\break}

\newwrite\efile \let\firsteqn=T
\def\writeqno#1%
{\if T\firsteqn \immediate\openout\efile=eqns.tmp\global\let\firsteqn=F\fi%
\immediate\write\efile{#1 \string#1}\global\advance\meqno by1}

\def\eqnn#1{\xdef #1{(\the\secno.\the\meqno)}\writeqno#1}
\def\eqna#1{\xdef #1##1{(\the\secno.\the\meqno##1)}\writeqno{#1{}}}

\def\eqn#1#2{\xdef #1{(\the\secno.\the\meqno)}%
$$#2\eqno(\the\secno.\the\meqno)$$\writeqno#1}
% ----------------------------->   Title page
\def\nobreak{\penalty1000}
\def\titl#1\endtitl{\par\vfil
     \vbox to 2in {}{\bf #1}\nobreak}
\def\titol#1\endtitol{\par\vfil
     \par\vbox to 1in {}{\bf #1}\par\vskip 1in\nobreak}
\def\tit#1\endtit{
     \vbox to 0.5in {}{\bf #1}\nobreak}

%
%
% math symbols
%
%---------------------------------------------------------------------
%
%
%
% space and backspace in l mode
%
\def\lspace{\ifx\answ\bigans{}\else\qquad\fi}
\def\lbspace{\ifx\answ\bigans{}\else\hskip-.2in\fi} % $$\lbspace...$$
%
%
%     curly letters
%
   %curly letters

%
%
%     derivatives
%
%

%

\def\bar#1{\overline{#1}}
\def\vev#1{\left\langle #1 \right\rangle}

\def\ket#1{\left| #1\right\rangle}
\def\abs#1{\left| #1\right|}

%
%
 %puts a small half in a displayed eqn
\def\frac#1#2{{\textstyle{#1\over #2}}} %puts a small fraction
%in a displayed eqn
%
%
%     various math operators
%
%
\def\tr{\mathop{\rm tr}}

\def\Im{\mathop{\rm Im}}
\def\kok{ {\rm K}^ 0-\overline{\rm K}{}^0 }
\def\bob{ {\rm B}^ 0-\overline{\rm B}{}^0 }

\def\eb{\epsilon_B}
\def\conj#1{\vbox{\ialign{##\cr
     $\scriptstyle{(}$--$\scriptstyle{)}$\cr\noalign{\kern-1pt\nointerlineskip}
     $\hfil\displaystyle{#1}\hfil$\cr}}}
\def\sconj#1{\vbox{\ialign{##\cr
     $\scriptscriptstyle{(}$--$\scriptscriptstyle{)}$\cr\noalign{\kern-1pt\nointerlineskip}
     $\hfil\scriptstyle{#1}\hfil$\cr}}}
\def\plus#1{\vbox{\ialign{##\cr
     $\scriptstyle{(}$+$\scriptstyle{)}$\cr\noalign{\kern-1pt\nointerlineskip}
     $\hfil\displayststyle{#1}\hfil$\cr}}}
\def\plus#1{\vbox{\ialign{##\cr
     $\scriptstyle{(}$+$\scriptstyle{)}$\cr\noalign{\kern-1pt\nointerlineskip}
     $\hfil\displaystyle{#1}\hfil$\cr}}}
\def\minus#1{\vbox{\ialign{##\cr
     $\scriptstyle{(}$--$\scriptstyle{)}$\cr\noalign{\kern-1pt\nointerlineskip}
     $\hfil\displaystyle{#1}\hfil$\cr}}}

\def\gl{ {\it \Gamma}_{\rm L} }

\def\bo{ {\rm B}^ 0 }
\def\bbar{ \overline{\rm B}{}^0 }
\def\bl{ {\rm B_{L}} }
\def\bh{ {\rm B_{H}} }

%
%
%       relations
%
\def\ltap{\ \raise.3ex\hbox{$<$\kern-.75em\lower1ex\hbox{$\sim$}}\ }
\def\gtap{\ \raise.3ex\hbox{$>$\kern-.75em\lower1ex\hbox{$\sim$}}\ }
\def\gl{\ \raise.5ex\hbox{$>$}\kern-.8em\lower.5ex\hbox{$<$}\ }
\def\roughly#1{\raise.3ex\hbox{$#1$\kern-.75em\lower1ex\hbox{$\sim$}}}
%
%
%       This defines et al., i.e., e.g., cf., etc.

\def\etal{\hbox{\it et al.}}
\def\[{\left[}
\def\]{\right]}
\def\({\left(}
\def\){\right)}
%
% some journal abbreviations
%

%
% Script Capital Macros
\textfont2=\tensy \scriptfont2=\sevensy \scriptscriptfont2=\fivesy
\def\cal{\fam2}
\def\Ascr{{\cal A}}
\def\Bscr{{\cal B}}

\def\Qscr{{\cal Q}}
\def\Rscr{{\cal R}}

\def\Uscr{{\cal U}}

%

% style abbreviations

\def\st{\scriptstyle}

% BOLDFACE Definitions
% From the TeXbook, a poor man's boldface in math mode:
\def\pmb#1{\setbox0=\hbox{$#1$}%
  \kern-.025em\copy0\kern-\wd0
  \kern.05em\copy0\kern-\wd0
  \kern-.025em\raise.0433em\box0}
% and reducing by a factor of .7 for \scriptstyle:
\def\pmbs#1{\setbox0=\hbox{$\st #1$}%
  \kern-.0175em\copy0\kern-\wd0
  \kern.035em\copy0\kern-\wd0
  \kern-.0175em\raise.0303em\box0}

\def\bfs#1{\hbox to .0035in{$\st#1$\hss}\hbox to .0035in{$\st#1$\hss}\st#1}
%\def\bfk#1{\hbox to .00385in{$#1$\hss}\textstyle #1}
%These are old definitions for boldface Greek displaystyle, scriptstyle 
% and kappa (which doesn't look good the first way).

% boldface greek

\def\bfGamma{\pmb{\Gamma}}

\def\Ttr{\pmb  \tau}
% miscellaneous boldface

\def\bfI{\hbox{\bf I}}

%
% Useful theoretical symbols
%%\newcommand{\brho}{\mbox{\boldmath $\rho$}}

\def\vev#1{\langle #1 \rangle}

\def\ket#1{\vert #1 \rangle }
%D'Alambertian 

%
%
\global\newcount\meqno \global\meqno=1
\newwrite\efile \let\firsteqn=T
\def\writeqno#1%
{\if T\firsteqn \immediate\openout\efile=eqns.tmp\global\let\firsteqn=F\fi%
\immediate\write\efile{#1 \string#1}\global\advance\meqno by1}

\def\eqqn#1#2{\xdef #1{(\the\meqno)}%
$$#2\eqno(\the\meqno)$$\writeqno#1}
%%%%%%%%%%%%%%%%%%%%%%%%%%%%%%%%%%
%\font\magonerm=cmr10 scaled \magstep1
%\font\magthreerm=cmr10 scaled \magstep3
%\font\magoneit=cmti10 scaled \magstep1
%\font\magonebf=cmb10 scaled \magstep1
%\font\maghalfrm=cmr10 scaled \magstephalf
\font\small=cmr9
%%\font\bigf=times at 30pt
%%\font\medf=times at 20pt
\font\medf=cmb10 scaled \magstep3
%%%%%%%%%%%%%%%%%%%%%%%%%%%%%%%%%%
%
\vsize=25 truecm
\hsize=16 truecm
\voffset=-0.8 truecm
%
%\oneandahalfspace
\singlespace
\parskip 6truept
\parindent 20truept
%
%%
%%
%\phantom{
\vbox{ {\rightline{\bf IFUM--FT 609/98}}
%       {\rightline{\bf UNIBAS--MATH X/97}}
}
%}
%%
%%
%
\hyphenation{ex-pe-ri-men-tal}
\hyphenation{va-cu-um}
\vskip 4truecm
\centerline{\medf 
                A Model--Independent and Rephase--Invariant}
\vskip 0.75truecm
\centerline{\medf
                Parametrization of CP--Violation}
\vskip 2truecm
\vskip 33truept
\centerline{D. Cocolicchio$^{(1,2)}$ and M. Viggiano$^{(1)}$}
\vskip 20truept
\vbox{
\centerline{\it $^{1)}$Dipartimento di Matematica,
Univ. Basilicata, Potenza, Italy}
\vskip 5truept
\centerline{\it Via N. Sauro 85, 85100 Potenza, Italy} }
\vskip 15truept
\vbox{
\centerline{\it $^{2)}$Istituto Nazionale di Fisica Nucleare,
                     Sezione di Milano, Italy}
\vskip 5truept
\centerline{\it Via G. Celoria 16, 20133 Milano, Italy} }
\vskip 3truecm
\centerline{\it ABSTRACT}
\vskip 15truept
\singlespace
\noindent
The phenomenological description of the neutral $B$ meson system is 
proposed in terms of the fundamental $CP$--violating observables and 
within a rephasing invariant formalism. \hfill\break
This generic formalism can 
select the time--dependent and time--integrated asymmetries which 
provide the basic tools to discriminate the different kinds of 
possible $CP$--violating effects in dedicated experimental $B$--meson 
facilities.
\vskip 1truecm
\noindent
{\sl 
}
\vskip 1truecm \noindent \singlespace
\vbox{
      {\leftline{PACS numbers: 12.15.Ff, 13.20.Jf, 14.40.Jz.
                              }}
      {\leftline{\it Keywords: $CP$--Violation.}}
%%      {\leftline{DECEMBER 1997}}
      }
\vfill\eject
\vsize=24 truecm
\hsize=16 truecm
\baselineskip 18 truept
\parindent=1cm
\parskip=8pt
\oneandahalfspace
%
%--------------------------------------------------------------------------
%
\phantom{.}
\blankline
\blankline
%
%%\leftline {\bf I. Introduction}
\blankline
\noindent
The time evolution of neutral meson decays can probably check
whether $CP$--violation arises from $CP$--violating phases in the
mixing (indirect $CP$--violation) or in the weak decay amplitudes
(direct $CP$--violation). Recently, there has been much interest
particularly on measuring the quark--mixing angles of the unitarity
triangle by means of the study of B--meson decays
\ref\Pavia{D. Cocolicchio,
``{\it $CP$--asymmetries in $B$ decays}'', Proc.
{\it Advanced Study Conference on Heavy Flavours},
ed. G. Bellini \etal\ (Ed. Frontieres, 1993), p. 367}.
Although, it has been observed
that there are some limitations in extracting these angles by using
the isospin ${\cal {SU}}(3)$ relations, nevertheless it is still 
tantalizing to
investigate carefully the time--dependence of correlative decay rates,
in order to extract the penguin effects into the $CP$--violating 
asymmetries for neutral $B$ meson decays. In this paper,
we propose a model--independent and rephase--invariant formalism
which provides the fundamental parameters directly in terms of the
measured quantities. The rephase--invariant formalism 
\ref\PalWu{
W. F. Palmer and Y. L. Wu, Phys. Lett. {\bf B350} (1995) 245 and refs. 
therein}\
will result more useful from
a phenomenological viewpoint, since it explicitly intends to
separate the different forms of $CP$--violation which a modern gauge
theory can account for. In a neutral meson system, a meaningful
classification of the three possible realizations of the $CP$--violating
asymmetries (through mixing, decay or mixing--and--decay) is
related to some phenomenological parameters which in principle
might be detected by studying the time evolution of the neutral
meson and some rate asymmetries.
The $\bob$ interfering effects into the $CP$--asymmetries can be
described in terms of an effective matrix Hamiltonian
\eqqn\hamilt{
H =
\cases{ M-{i \over 2}\Gamma\qquad & \ref\uno{
L. Lavoura, Ann. Phys. (NY) {\bf 207} (1991) 428;
C. D. Buchanan, R. Cousin, C. Dib, R. Peccei and J. Quackenbush,
Phys. Rev. {\bf D45} (1992) 4088}
 \cr
& \cr  
{\cal H}_\mu \sigma_\mu = (E_1 \sigma_1 + E_2 \sigma_2 
E_3 \sigma_3) -  i D {\bf 1} \qquad &
\ref\due{M. Hayakawa and A. I. Sanda, Phys. Rev. {\bf D48} (1993) 
1150; M. Kobayashi and A. I. Sanda, Phys. Rev. Lett. {\bf 69} (1992) 
3139; see also D. Cocolicchio, L. Telesca and M. Viggiano, e-preprint archive 
hep-ph/9709486}
\cr 
& \cr
-i \left(
\matrix{
 i d & b^2 \cr
b^{\prime 2} & i{d^\prime} 
\cr}
\right) 
\qquad & \ref\tre{S. H. Aronson, G. J. Bock, H.-Y. Cheng and E. Fishbach,
Phys. Rev. {\bf D28} (1983) 495} \qquad . 
\cr}
}
In Table I, we propose the transformation properties 
of the elements of the Hamiltonians,
with respect to some important combinations like $CP$ and $CPT$ 
of the discrete space--time symmetries.

\bigskipamount=30pt plus2pt minus2pt
%_______________________________________________________________________________
%
\phantom{.}
\blankblankline
%_______________________________________________________________________________
%
\midinsert{
\centerline{\vbox{
\hsize=17truecm
\singlespace
\noindent
%\centerline{
{\bf TABLE I.} 
{\small
The restrictions imposed by combinations of charge conjugation $C$, 
parity $P$ and, time reversal $T$ on the elements of the Hamiltonian 
matrix.}
%}
\vskip 7pt
\centerline{\vbox{\tabskip=0pt\offinterlineskip
\def\tablerule{\noalign{\hrule}}
\def\vsptwelve{height12pt&\omit&&&\omit&}
\halign to \hsize{
\vrule# 
\tabskip=12pt plus2pt minus8pt
& $\;$ #\hfil 
& $\;$ \hfil#\hfil 
& $\;$ \hfil#\hfil
& $\;$ \hfil#\hfil
\tabskip=12pt plus2pt minus8pt
& \vrule#\tabskip=0pt\cr
\tablerule
\vsptwelve\cr
& Form of H 
& $CPT$ 
& $T$ 
& $CP$  & \cr
height12pt&\omit&&&\omit&\cr
\tablerule
height14pt&\omit&&&\omit&\cr
&&&&$\phantom{OO}$&\cr
&          
& $M_{11}=M_{22}$ 
& $M_{12}=M_{12}^\ast = M_{21} =M_{21}^\ast $
& $M_{12}=M_{12}^\ast=M_{21}=M_{21}^\ast$                & \cr
height18pt&\omit&&&\omit&\cr
&&&&$\phantom{OO}$&\cr
& $M-{i\over 2} \Gamma$ 
& $\Gamma_{11}=\Gamma_{22}$ 
& $\Gamma_{12}=\Gamma_{12}^\ast = \Gamma_{21}=\Gamma_{21}^\ast $
& $\Gamma_{12}=\Gamma_{12}^\ast=\Gamma_{21}=\Gamma_{21}^\ast$  & \cr
height18pt&\omit&&&\omit&\cr
&&&&$\phantom{OO}$&\cr
&
& 
& 
& $M_{11}=M_{22}^\ast \, , \quad \Gamma_{11}=\Gamma_{22}^\ast$     & \cr
height18pt&\omit&&&\omit&\cr
&&&&$\phantom{OO}$&\cr
height18pt&\omit&&&\omit&\cr
&&&&$\phantom{OO}$&\cr
& ${\cal H}_\mu\sigma_\mu $ 
& ${\cal H}_z=0 $ 
& ${\cal H}_y =0 $ 
& ${\cal H}_y={\cal H}_z=0 $                                   & \cr
height18pt&\omit&&&\omit&\cr
&&&&$\phantom{OO}$&\cr
height18pt&\omit&&&\omit&\cr
&&&&$\phantom{OO}$&\cr
& $d,d^\prime,b,b^\prime$ 
& $d=d^\prime$ 
& $b^2=b^{\prime 2}$ 
& $d=d^\prime, \quad b^2=b^{\prime 2}$  & \cr
height18pt&\omit&&&\omit&\cr
&&&&$\phantom{OO}$&\cr
%height12pt&\omit&&&&&\omit&\cr
\tablerule\hfil\cr}}}
}}}
\endinsert
%
%------------------------

\noindent
The effective Hamiltonian matrix is determined by eight real 
parameters, but only seven are physical meaningful because the 
absolute phase of $H_{12}$ or $H_{21}$ is meaningless, being the 
relative phase of $\bo$ and $\bbar$ arbitrary.
They can be substituted by the two complex eigenvalues
\eqqn\eIIad{
\lambda_{H,L} = {1\over 2} \left( C \mp D \right) }
where, in a general theory,
$C =  \lambda_L +\lambda_H = H_{11} + H_{22} =\tr H $ and
$D^2 = (\lambda_L -\lambda_H)^2 = \left(H_{11} - H_{22}\right)^2 
+ 4 H_{12} H_{21} = (\tr H)^2 - 4 (\det H)$,
and by the two complex mixing parameters $\epsilon_{H, L}$
which are given by:
\eqqn\eIIepsl{
\eqalign{
\epsilon_H =& 
\left({ {2 H_{12} - D}
             \over
        {2 H_{12} + D} }\right)  -
\left({ {4 H_{12}}
              \over
         {2 H_{12} + D} }\right) 
\left({  {H_{11} - H_{22}}
             \over
         {H_{11} - H_{22} + D + 2 H_{12} } }\right) = \eb - \delta_H \cr
\epsilon_L =& 
\left({ {2 H_{12} - D}
             \over
         {2 H_{12} + D} }\right)  -
\left({ {4 H_{12}}
              \over
         {2 H_{12} + D} }\right) 
\left({  {H_{11} - H_{22}}
             \over
         {H_{11} - H_{22} - D - 2 H_{12}} }\right) = \eb - \delta_L \; , \cr}
}
where

\eqqn\eIIeps{
\eqalign{
\eb = &  { {2 H_{12} - D}
              \over
               { 2 H_{12} + D}   } 
          = {  { \sqrt{H_{12}} - \sqrt{H_{21}} } 
                           \over
               { \sqrt{H_{12}} + \sqrt{H_{21}} } } 
\cr
\delta_H =& 
\left({ { 2 H_{12}}\over {2 H_{12} + D} }\right)
\left({ {H_{11} - H_{22} }
                                                    \over
{H_{11} + H_{12} -\lambda_H} }\right)
\cr
\delta_L =& 
\left({ { 2 H_{12}}\over {2 H_{12} + D} } \right)
\left({ {H_{11} - H_{22} }
                                                    \over
        {H_{11} - H_{12} -\lambda_L} }\right) \; .
\cr} }
The complex scaling matrix $\Rscr$, which
diagonalizes the effective Hamiltonian,
is then given by
\eqqn\eIIau{
\Rscr = 
\left(\matrix{
N_H (1+\epsilon_H)  & -N_H (1-\epsilon_H) \cr
N_L (1+\epsilon_L) & \phantom{-}N_L(1-\epsilon_L)\cr}\right)
= \left(\matrix{ p & - q \cr
                 p^\prime  & q^\prime \cr}\right)
\, ,
}
where $N_{H,L}^{-2}= 2(1+|\epsilon_{H,L} |^2)$ are fixed only once the 
eigenvectors
normalization is realized and
\eqqn\eIIeta{
\eqalign{
\eta_H =&{ {- q}\over p} = - { {(1-\epsilon_H)}
                  \over
             {(1+\epsilon_H)} }
       = - { {(H_{11} - H_{22} + D)}
                    \over
              {2 H_{12}} }
       = \phantom{-}{    {2 H_{21}}
                 \over
           {(H_{11} - H_{22} - D)} } \cr
\eta_L =& { { q^\prime}\over {p^\prime}} =\phantom{-} { {(1-\epsilon_L)}
                  \over
             {(1+\epsilon_L)} }
       = - { {(H_{11} - H_{22} - D)}
                    \over
              {2 H_{12} } }
       = {    {2 H_{21}}
                 \over
           { (H_{11} - H_{22} + D) } }  \; . \cr }
}
In any $CPT$--invariant theory, we consider from now onward,
$H_{11} = H_{22}$ and there is only one mixing parameter 
$\eb=\epsilon_H=\epsilon_L$ and therefore $\eta_H = -\eta_L$,
$N_H^{-2}=N_L^{-2}=N^{-2}=2(1+|\eb |^2)$, $p=p^\prime$, $q=q^\prime$.
In this case, we have that
\eqqn\eIIaad{
\lambda_H = H_{11} -\sqrt {H_{12}H_{21}}=M_{11}-{i\over 2} \Gamma_{11}
-{D\over 2}=m_H -{i\over 2} \gamma_H
}
\eqqn\eIIaat{
\lambda_L = H_{11} +\sqrt {H_{12}H_{21}}=M_{11}-{i\over 2} \Gamma_{11}
+{D\over 2}=m_L -{i\over 2} \gamma_L 
}
with
\eqqn\eIIaaq{
D = 2 \sqrt{ H_{12} H_{21} } = 2
\sqrt{(M_{12}-{i\over 2} \Gamma_{12}) ({M_{12}}^*-{i\over 2} 
{\Gamma_{12}}^*)}=-\left(\Delta m - {i\over 2} \Delta\gamma\right)\; , }
being $\Delta m = m_H - m_L$ and $\Delta \gamma = \gamma_H - \gamma_L$.
These real ($m_{H,L}$) and imaginary ($\gamma_{H,L}$)
components will define the 
masses and the decay widths of the eigenstates $\bh$ and $\bl$
in the narrow width approximation. 
These heavy and light particles are then a linear combination 
of the flavour $\bo$ and $\bbar$ states:
\eqqn\eIIo{
\left( \matrix{ \vert \bh \rangle \cr 
            \cr
          \vert \bl \rangle \cr}\right) =
\Rscr
\left( \matrix{ \vert \bo \rangle \cr 
     \cr     
     \vert \bbar \rangle \cr}\right) \; ,
}
where usually $\Rscr$ is preferably parameterized
according to the following relations
\eqqn\EDFGH{
\Rscr = 
\cases{ 
 {\displaystyle {1\over {\sqrt{2(1+\vert\eb\vert^2)}}}} 
\left(
\matrix{
(1+\eb) & -(1-\eb) \cr
(1+\eb) &  (1-\eb) \cr}
\right) \cr
\cr \cr
{\displaystyle {\vert 1 - \eta \vert \over 1 - \eta}}  
{\displaystyle {1\over{\sqrt{1+\vert\eta\vert^2}}}}
\left( \matrix{ 1 & \eta \cr
         1 & -\eta \cr}
\right) \cr
\cr \cr
\left(
\matrix{
p & -q \cr
p & q \cr}
\right) \; . \cr}
}
After the corresponding normalization of the eigenvectors
\eqqn\eIIn{
\langle \bl \vert \bl \rangle = \langle \bh \vert \bh \rangle
= |p|^2 + |q|^2 = 1\; ,
}
the impurity parameters can be connected by the simple relations
\eqqn\eIIx{
\eb = \frac{\displaystyle p-q }{\displaystyle p+q} 
={  { \displaystyle (\sqrt{H_{12}} - \sqrt{H_{21}}) }
                  \over
    {\displaystyle (\sqrt{H_{12}} + \sqrt{H_{21}}) } }
= i { {\displaystyle \Im M_{12} - {i\over 2} \Im \Gamma_{12} }
                  \over
      {\displaystyle {\rm Re} \, M_{12} - {i\over 2} {\rm Re}\, 
\Gamma_{12} + {D\over 2} } }
\; .
}
We want to stress the crucial issue that the phase of $p$ and $q$ 
may be altered by redefining
the relative phase between the state vectors $\ket\bo$ and $\ket\bbar$.
Therefore both $p$ and $q$ are {\it not} measurable quantities. 
Thus, it results evident that the $CP$-violation parameter $\epsilon_B$
arises from a relative imaginary part between the
off-diagonal elements $M_{12}$ and $\Gamma_{12}$ i.e. if 
$\delta = arg(M_{12}\Gamma_{12}^*)\ne 0$.
To this end, we introduce the ratio between these relevant variables
\eqqn\ratio{
{{M_{12}}\over{\Gamma_{12}}}=
{ {\vert{M_{12}\vert} \over {\vert{\Gamma_{12}}\vert} } }e^{i\delta}
= - r e^{i \delta} ,
}
where the relative phase is $\delta = (\delta_M - \delta_\Gamma)$.
In terms of $r$ and $\delta$, the $CP$--violating parameter $\epsilon_B$ 
is obtained as follows
\eqqn\erdelt{
\epsilon_B \simeq i {\sin \delta_M \over 1 + \cos \delta_M} +
\left( {1 \over 1 + \cos \delta_M} \right) \, 
\left({2r - i \over 4r^2 + 1} \right) \delta.
} 
To the extent that 
\eqqn\eext{
\eqalign{
M_{12} = & \vert M_{12} \vert e^{ i \delta_M } \cr
\Gamma_{12} = & \vert \Gamma_{12} \vert e^{i \delta_\Gamma } \cr}
}
share the same phase $\delta_M = \phi = \delta_\Gamma$,
we have no $CP$--violation and we obtain that 
\eqqn\FRTG{
\epsilon_B = i { \vert M_{12} \vert \sin \delta _M - {\displaystyle {i \over 2}} 
\vert \Gamma_{12} \vert \sin \delta _\Gamma \over \vert M_{12} \vert 
\cos \delta _M - {\displaystyle {i \over 2}}
\vert \Gamma_{12} \vert \cos \delta _\Gamma +{\displaystyle {D \over 2}}}
= i {\sin \phi \over 1 + \cos \phi}
} 
independently of any phase convention.
However, it is clear that only the magnitude of
\eqqn\magn{
\eta = -{q \over p} = - {1-\eb \over 1+\eb} = - \sqrt{M^\ast_{12} -
{i \over 2}\Gamma^\ast_{12} \over M_{12} -
{i \over 2}\Gamma_{12}}
}
results a measurable quantity and it results connected to the 
following overlap parameter
\eqqn\overlap{
s = \vev {\bh \vert \bl} = 
{ {2 {\rm Re} \, \eb}
     \over
  {1 + |\eb|^2} } 
=  { {1 -\vert\eta\vert^2}
      \over
   {1+\vert\eta\vert^2} }
\simeq { {2r}\over{4r^2 +1}}\delta  
\; .
}
This means that $CP$--nonconservation is determined by the relative 
phase between $M_{12}$ and $\Gamma_{12}$.
The value of the
parameter $\vert \eta \vert$ is significant,
in the sense that $\eta = {1-\eb \over 1+\eb}
\neq 1$ does not necessarily imply $CP$--violation.
$CP$ is violated in the mixing matrix
if $\vert \eta\vert \ne 1$.
Remember that, since flavour is conserved in strong interactions, 
there is some freedom in defining the phases of flavour eigenstates. 
This means that $\eta$ is a phase dependent quantity 
manifesting its presence in the phase of $\epsilon_B$ which must only 
satisfy Eq. \overlap .
In turn, Eq. \overlap\ reduces to the equation of a circumference
\eqqn\overuno{
\left( \hbox{Re}\epsilon_B - {1\over s}\right)^2 +\left(\hbox{Im}\epsilon_B
\right)^2 = \left({1\over 
s}\right)^2 -1
}
of radius $\sqrt{({1\over s})^2 -1}\simeq {1\over s}- {s \over 2} $
centered in ($1\over s$, 0) of the Gauss complex $\epsilon_B$-plane.
This geometric interpretation of the dependence of $\epsilon_B$ on the
choice of the phase convention means that $\epsilon_B$ picks
a point on the circumference of this circle according to each possible 
convention. 
The relative pure phase $\delta$ can be derived from the fact that
\eqqn\overdue{
\eta \simeq
-e^{i \delta_\Gamma}
\left [1 - { {2r}\over {4r^2 +1}} ( 1+ 2 i r)\delta \right ].
}
Assuming the 
$\Delta\Bscr = \Delta \Qscr$
rule conserved
\ref\DBDQ{
G. V. Dass and K. V. L. Sarma, Phys. Rev. {\bf D54} (1996) 5880},
the amount of its magnitude can then be extracted 
from the decay rate asymmetry between the semileptonic channels
$B \rightarrow \ell^+\nu_\ell X^-$ and ${\overline B} \rightarrow 
\ell^-{\overline \nu}_\ell X^+$. The related $CP$--violating 
asymmetry
\eqqn\esl{
A^{\ell \ell}_{SL} = 
{{1-|\eta|^4}\over{1+|\eta|^4}}
= {4r\sin \delta \over 4r^2 + 1}
 \; ,
}
has been predicted theoretically 
\ref\cd{D. Cocolicchio and L. Maiani, Phys. Lett. {\bf B291} (1992) 155}
%T. Altomari, L. Wolfenstein and J. P. Bjorken, Phys. Rev. {\bf D 37} (1988) 1800;
%M. Lusignoli, Z. Phys. {\bf C 41} (1989) 645}
\ and somewhat
detected experimentally \ref\cleo{CLEO Collaboration, J. Bartelt \etal,
Phys. Rev. Lett. {\bf 71} (1993) 1680; CDF Collaboration, F. Abe \etal,
Phys. Rev. {\bf D55} (1997) 2546; OPAL Collaboration, K. Ackerstaff \etal,
Z. Phys. {\bf C76} (1997) 401}.
However, it is worth noting that the experimental results \cleo \ of 
the decay rate asymmetry $A^{\ell \ell}_{SL} $ do not constrain $\epsilon_B$ 
in a sensible way, due to the lack of available data of other direct 
$CP$--violating effects. Analogously to the case of the neutral kaon 
mixing, the magnitude and the phase of the complex parameter 
$\epsilon_B$ depend on the specific phase--convention
adopted to describe the $\bob$ system. Indeed, 
this phase--convention dependence could induce errors to provide the 
experimental information on $\epsilon_B$
\ref\ddwu{
L. Lavoura, Mod. Phys. Lett. {\bf A7} (1992) 1367; 
A. I. Sanda and Zhi-zhong Xing, Phys. Rev. {\bf D56} (1997) 6866;
Dan-Di Wu, Prairie View A \& M Univ. preprint PVAM-HEP-9-97;
M. C. Banuls and J. Bernabeu, preprint CERN-TH/97-216}. 
For a particular phase--convention, $M_{12}$ may turn out to have
a large phase, but without also knowing the phase of $\Gamma_{12}$, no
conclusion can be reached as to the size of indirect $CP$--nonconservation.
In the Wu-Yang convention $\hbox{Im}\Gamma_{12}=0$, we obtain that
\eqqn\dsw{
\arg(\epsilon_B)\simeq\cases{
\pi +\Phi_{SW}\quad\hbox{for}\quad\hbox{Im}{\epsilon_B}>0 \;,\cr
\Phi_{SW}\quad\hbox{for}\quad\hbox{Im}{\epsilon_B} < 0 \;,\cr}
}
being the superweak phase $\Phi_{SW}=\tan^{-1}(2 r)$.
The question we refer here arises 
whatever we introduce an a priori assumption for a phase convention 
dependent observable, like in the case of $\epsilon_B$.
Assuming the box diagram dominance of the Standard Model,
the situation is still unclear, being
${ {\vert M^B_{12} \vert} \over {\vert \Gamma^B_{12} \vert} }$
very large and
$\vert \eta\vert\simeq 1$.
From the available experimental inputs of
$x_d\simeq \Delta m_B\tau_B $ and
$\Delta m_B\simeq 2  \vert M^B_{12}\vert $,
we find that 
$\delta \simeq \left( {{\Im M^B_{12}}\over {\vert 
M^B_{12}\vert}}\right)$ is in a wide range.
Consequently, we may say that the rephasing dependence of the impurity 
parameter $\epsilon_B$ is of no use for the study of the $CP$--violating
effects in the $\bob$ system.
This example suggests to develop 
a formalism more transparent from a 
phenomenological point of view and 
expressible directly in terms of observables.
Therefore, a suitable choice of the relative phase between $CP \ket \bo = 
e^{i \delta_{CP}}\ket \bbar$
has to be adopted for the specification of the parameter $\eb$.
Usually, we fix this relative phase between the two states and then
we determine the $CP$--nonconserving effects from the relative
phase between $M_{12}$ and $\Gamma_{12}$. 
But, as we stressed before,
this approach can induce some ambiguities if not errors and, 
therefore, a more general formalism is needed. In order to develop a 
generalized rephasing invariant method we cannot forget that,
although the properties of the particle mixing 
are connected to the solution of
a Schr\"odinger equation of an effective Hamiltonian $H = M - {i \over 2}
\Gamma$, nevertheless the essential tool of the description
is represented by the transition amplitude that is given in matrix notation by
\eqqn\btran{
{\Ttr}_{FI} = \sum_{ij} A_{Fi}^\ast\big[ s\bfI - H \big]
_{ij}^{-1} A_{jI}=
{\pmb{A}}_D^{\dag}\big[ s\bfI - H \big]^{-1}
{\pmb{A}}_P
}
in terms of the production ${\pmb{A}}_P = (A_{iI})$ and decay
${\pmb{A}}_D = (A_{jF})$ amplitudes of the production vertex $P$ of the initial 
$I$ channel mode and the decay vertex $D$ of the final $F$--decay mode.
An initial B--meson in a pure state decays into an $F$--channel mode
with an amplitude
\eqqn\amplit{
A_{iF}(t) = \left\langle {F} \vert {B_i (t)}\right\rangle =
\left\langle F \vert {\cal U}_{ij}(t) \vert B_j \right\rangle 
}
being 
\eqqn\bein{
A_D = (A_{iF}) = 
\left(\matrix{\left\langle {F} \vert {\bo}\right\rangle \cr
\cr
                   \left\langle {F} \vert {\bbar}\right\rangle \cr}\right) =
\left(\matrix{A (\bo \rightarrow F) \cr
\cr
              A (\bbar \rightarrow F) \cr}\right) 
}
and 
\eqqn\eIIevol{
\Uscr (t) = 
\left(
\matrix{
g_1 (t) &  \eta g_2 (t) \cr
& \cr
{\displaystyle {1\over\eta}} g_2 (t) & g_1 (t) \cr}\right) 
}
written in terms of
$g_{1,2} (t) = 
{1\over 2} (\hbox{e}^{- {\rm i} \lambda_{H} t} \pm
   \hbox{e}^{ - {\rm i} \lambda_{L} t} ) $
\ref\AcutoCD{A. Acuto and D. Cocolicchio, Phys. Rev. {\bf D47} (1993) 3945}.
If ${\overline F}$ denotes the charge--conjugate final decay state, we have
the conjugate amplitude
\eqqn\conjug{
A_{i{\overline F}} = \vev {{\overline F} \vert {\rm B}_i (t)} \quad
{\rm and} \quad {\overline A}_D = (A_{i{\overline F}}) \;.
}
Squaring, we obtain the time--dependent rates
\eqqn\tdep{
\bfGamma ({\rm B}_i (t) \rightarrow {\conj F}) = 
\sum^{}_{kj} {\cal U}^{\ast}_{
ki}{\cal U}_{kj} \vert A_{j{\sconj F}} \vert ^2
}
where the summation over repeated indices is tacitly assumed.
The time--integrated rates of interest
\eqqn\widhts{
{\widehat {\bfGamma}}({\rm B}_i \rightarrow {\conj F}) = 
\int_0^\infty dt \, \bfGamma({\rm B}_i (t) \rightarrow {\conj F})=
\int_0^\infty dt \,
\vert A_{i{\sconj F}} (t) \vert ^2 \; ,
}
can be expressed in terms of the rephase--invariant complex parameters
\eqqn\Cmplprm{
{\conj \xi}_F\!=\!{ q \over  p}
{ A  (\bbar\rightarrow {\conj F}) \over
A (\bo\rightarrow {\conj F})} =\!{ q \over  p}
{ {\overline A}_{\sconj F} \over
A_{\sconj F}} \; , 
}
with the following resulting expressions
\eqqn\becom{
\eqalign{
 &{\widehat {\bfGamma}}(\bo \rightarrow {\conj F}) = 
\vert A_{\sconj F} \vert^2 \Big[{\widehat{\cal M}}_{11}  + \vert {\conj \xi} _F 
\vert ^2
{\widehat{\cal M}}_{22} - 2 {\rm Re} \Big( {\conj \xi}_F {\widehat{\cal M}}
_{21}\Big)\Big] \cr
& {\widehat {\bfGamma}}(\bbar \rightarrow {\conj F}) = 
\vert A_{\sconj F} \vert^2 \Big[{\widehat{\cal M}}_{22}  + 
\vert {\conj \xi} _F \vert ^2
{\widehat{\cal M}}_{11} - 2 {\rm Re} \Big( {\conj \xi}_F {\widehat{\cal M}}
_{12}\Big)\Big]
\vert \eta \vert ^{-2} \; ,
\cr }
}
being ${\widehat{\cal M}}_{ij} = \int_0^\infty dt \, 
{\cal M}_{ij} (t)$ with ${\cal M}_{ij} = g_i g^\ast_j$ \AcutoCD \
and, supplemented by the unitarity sum rule of Bell and Steinberger
\ref\BS{
J. S. Bell and J. Steinberger, Proceedings Oxford Int. 
Conf. on Elementary Particles 1965, ed. R. G. Moorehouse \etal\
(Rutherford HEP Lab., Chilton, Didcot, Berkshire, England, 1966) p. 
195},
\eqqn\SRBS{
\eqalign{
\langle \bh \vert \bfGamma \vert \bl \rangle = &
\Big[ \big( {{\gamma_H +\gamma_L}\over 2} \big) - i \big( 
m_H-m_L\big)\Big] \left\langle {\bh} \vert {\bl}\right\rangle = \cr
= & \sum_F \langle F\vert H \vert \bh \rangle^\ast 
\langle F\vert H \vert \bl \rangle = 
\sum_F \langle \Gamma_F\rangle \big( \Ascr + i \Bscr) \cr}
}
where the last expression has been obtained by choosing the final 
decay modes $F$ to be $CP$--eigenstates and the integration
with respect to the phase space must be understood. We have defined
\eqqn\Fgam{
\langle \Gamma_F\rangle  =
{1 \over 1+\vert \eta \vert ^2} \Big[ \bfGamma(\bo\rightarrow F)
+  \vert \eta \vert ^2 \bfGamma(\bbar\rightarrow F) \Big] \simeq
{1\over 2}
\Big[ \bfGamma(\bo\rightarrow F) +
\bfGamma(\bbar\rightarrow F)\Big]
\; .}
The two independent $CP$--violating parameters are given by
\eqqn\AeB{
\eqalign{
\Ascr=& {{1-\abs{\xi_F}^2} \over {1+\abs{\xi_F}^2} } \cr
\Bscr=& {{2\Im\xi_F}\over{1+\abs{\xi_F}^2} } \; , \cr}
}
and
$\langle {\rm B}_L\vert {\rm B}_H \rangle
={ { 1- \vert \eta \vert ^2 } \over {1+
\vert \eta \vert ^2 }}\simeq 10^{-3}$
imposes large cancellations in the sum.

\noindent
In the most general case of a state ${\overline F}$ which is not $CP$ 
self--conjugate of the final decay mode $F$
\ref\BiSa{
H. J. Lipkin, Y. Nir, H. R. Quinn and A. E. Snyder, Phys. Rev.
{\bf D44} (1991) 1454}, we can introduce
the following relevant parameters
\eqqn\relev{
\eqalign{
\epsilon^\prime_{ij} & = {1 - f_{ij} \over 1 + f_{ij}} \cr
\epsilon^{\prime \prime}_{ij} & = {1 - {\overline f}_{ij} \over 1 + 
{\overline f}_{ij}} \cr
\epsilon^{CP}_{ij} & = {1 - f^{CP}_{ij} \over 1 + f^{CP}_{ij}} \; , \cr}
}
being
\eqqn\fffb{
f_{ij} = {A_{iF} \over A_{jF}} \, , \quad 
{\overline f}_{ij} = {A_{i{\overline F}} \over A_{j{\overline F}}} 
\quad {\rm and} \quad 
f^{CP}_{ij} = {A_{iF} \over A_{j{\overline F}}} \; ,
}
or alternatively we can introduce the parameters
\eqqn\phystate{
\eta_{\alpha \beta} = {A_{\alpha F} \over A_{\beta F}} =
{R_{\alpha i} f_{ij} + R_{\alpha j} \over R_{\beta i} f_{ij} + R_{\beta j}}
\qquad {\rm and } \qquad 
{\overline \eta}_{\alpha \beta} = {A_{\alpha {\overline F}} \over 
A_{\beta {\overline F}}} =
{R_{\alpha i} {\overline f}_{ij} + R_{\alpha j} \over 
R_{\beta i} {\overline f}_{ij} + R_{\beta j}} \; .
}
The above parameters $f_{ij}$ and ${\overline f}_{ij}$ can describe 
the direct $CP$--violating effects,
but indeed, they are not physical observables since they are not 
rephase--invariant. 
Indeed, the relevant phase--independent observables are represented only by
\eqqn\relreph{
\eqalign{
\eta_{F} = & {\left\langle {F} \vert B_H \right\rangle \over
\left\langle {F} \vert B_L \right\rangle} = {1 - \xi_F \over 
1 + \xi_F} \cr
{\overline \eta}_F = & {\left\langle {\overline F} \vert B_H 
\right\rangle \over
\left\langle {\overline F} \vert B_L \right\rangle} = {1 - {\overline \xi}_F 
\over 
1 + {\overline \xi}_F} . \cr}
}
Physically, we can introduce the following possible rate asymmetries
\eqqn\asymm{
\eqalign{
a_{ij} = & {\vert A_{iF} (t) \vert^2 - \vert A_{jF} (t) \vert^2 \over 
\vert A_{iF} (t) \vert^2 + \vert A_{jF} (t) \vert^2} = 
-{2 {\rm Re} \ \epsilon^\prime_{ij} \over 1 + \vert \epsilon^\prime_{ij} \vert^2} 
\cr
{\overline a}_{ij} = & {\vert A_{iF} (t) \vert^2 - \vert A_{j{\overline F}} (t) 
\vert^2 \over 
\vert A_{iF} (t) \vert^2 + \vert A_{j{\overline F}} (t) \vert^2} =
-{2 {\rm Re} \ \epsilon^{CP}_{ij} \over 1 + 
\vert \epsilon^{CP}_{ij} \vert^2}
%{\left \vert f_{ij} {{\displaystyle 
%A_{jF}}\over {\displaystyle A_{j{\overline F}}}} 
%\right\vert^2 -1
%\over {\left \vert f_{ij} {{\displaystyle 
%A_{jF}}\over {\displaystyle A_{j{\overline F}}}} 
%\right\vert^2 +1 }
%}
\cr
{\overline {\overline a}}_{ij} = & {\vert A_{i{\overline F}} (t) \vert^2 - 
\vert A_{j{\overline F}} (t) 
\vert^2 \over 
\vert A_{i{\overline F}} (t) \vert^2 + \vert A_{j{\overline F}} (t) \vert^2} = 
-{2 {\rm Re} \ \epsilon^{\prime \prime}_{ij} \over 1 + 
\vert \epsilon^{\prime \prime}_{ij} \vert^2} \; .
\cr}
}
In principle, these rephase--invariant quantities
represent the solution of the problem. To be more specific
and show a link with previous results \PalWu,
we can define these asymmetries by means of the following four time--dependent
decay rates
\eqqn\sevasym{
\Gamma (\bo (t) \rightarrow {\conj F}) = 
\cases{
{\displaystyle{1 \over 2}} \vert A_{\sconj F} \vert^2
e^{-\Gamma t} \biggl[ {\conj C}_y \cosh (y \Gamma t)  
+{\conj C}_x \cos (x \Gamma t) + \cr
\phantom{{1 \over 2} \vert A \vert^2
e^{-\Gamma t} }
+ {\conj S}_y \sinh (y \Gamma t) + {\conj S}_x \sin (x \Gamma t) \biggr] \cr
\cr \cr
{\displaystyle{1 \over 4}}\vert A_{\sconj F} \vert^2 \biggl [ f_1(t) \vert {\conj \xi}_F \vert^2 + 
f_2(t) + 2f_3(t){\rm Re} {\conj \xi}_F + 2f_4(t) {\rm Im} {\conj \xi}_F \biggr] \cr
\cr  \cr
{\displaystyle{1 \over 4}}\vert A_{\sconj F} \vert^2 \biggl [ h_1(t) \vert 1 - {\conj \xi}_F \vert^2 + 
h_2(t) \vert 1 + {\conj \xi}_F \vert^2 + \cr
\phantom{.}
+ h_3(t){\rm Re} \big [(1 + {\conj \xi}_F)(1 - {\conj {\xi^\ast}}_F)\big ] + 
h_4(t) {\rm Im} \big [(1 + {\conj \xi}_F)(1 - {\conj {\xi^\ast}}_F) \big ] \biggr]
 \cr}
}
and 
\eqqn\sevasymII{
\Gamma (\bbar (t) \rightarrow {\conj F}) = 
\cases{
{\displaystyle{1 \over 2}} \vert A_{\sconj F} \vert^2
e^{-\Gamma t} \biggl[ {\conj {C^\prime}}_y \cosh (y \Gamma t)  
+{\conj{C^\prime}}_x \cos (x \Gamma t) + \cr
\phantom{{1 \over 2} \vert A \vert^2
e^{-\Gamma t} }
+ {\conj {S^\prime}}_y \sinh (y \Gamma t) + 
{\conj {S^\prime}}_x \sin (x \Gamma t) \biggr] \cr
\cr \cr
{\displaystyle{1 \over 4}}
{\displaystyle{\vert A_{\sconj F} \vert^2 \over \vert \eta \vert^2}} 
\biggl [ f^\prime_1(t) \vert {\conj \xi}_F \vert^2 + 
f^\prime_2(t) + 2f^\prime_3(t){\rm Re} {\conj \xi}_F + 2f^\prime_4(t) {\rm Im} {\conj \xi}_F \biggr] \cr
\cr  \cr
{\displaystyle{1 \over 4}}
{\displaystyle{\vert A_{\sconj F} \vert^2 \over \vert \eta \vert^2}}  
\biggl [ h^\prime_1(t) \vert 1 - {\conj \xi}_F \vert^2 + 
h^\prime_2(t) \vert 1 + {\conj \xi}_F \vert^2 + \cr
\phantom{.}
+ h^\prime_3(t){\rm Re} \big [(1 + {\conj \xi}_F)(1 - {\conj {\xi^\ast}}_F) \big ] + 
h^\prime_4(t) {\rm Im} \big [(1 + {\conj \xi}_F)(1 - {\conj {\xi^\ast}}_F) \big ] \biggr]
 \cr}
}
where the time--dependent functions are defined by
\eqqn\fghdef{
\vcenter{\openup1\jot \halign{$\hfil#$&${}#\hfil$&
\quad$\hfil#$&${}#\hfil$\cr
f^{(\prime)}_1(t)&=e^{-\gamma _L t}+e^{-\gamma _H t}{\plus-}
2e^{- \Gamma t}\cos (\Delta m t) ,
&f^{(\prime)}_2(t)&=e^{-\gamma _L t}+e^{-\gamma _H t}{\minus +}
2e^{- \Gamma t}\cos (\Delta m t),\cr
f^{(\prime)}_3(t)&=e^{-\gamma _L t}-e^{-\gamma _H t},&f^{(\prime)}_4(t)
&={\plus -}2 e^{- \Gamma t}\sin (\Delta m t)\, ;\cr
& & & \cr
h^{(\prime)}_1(t)&=
%{\displaystyle {f^{(\prime)}_1(t) + f^{(\prime)}_2(t) - 
%2f^{(\prime)}_3(t) \over 4}}
e^{-\gamma_H t},&h^{(\prime)}_2(t)&=
%{\displaystyle 
%{f^{(\prime)}_1(t) + f^{(\prime)}_2(t) + 2f^{(\prime)}_3(t) \over 4}}
e^{-\gamma_L t},\cr
h^{(\prime)}_3(t)&=
%{\displaystyle -{f^{(\prime)}_1(t) - f^{(\prime)}_2(t) \over 2}}
{\minus +}2e^{-\Gamma t} \cos (\Delta mt)\, ,
&h^{(\prime)}_4(t)&=
%f^{(\prime)}_4(t) 
{\plus -}2e^{-\Gamma t} \sin (\Delta mt) \, .
\cr}}
}
Here, we have preferred to use the following two dimensionless
parameters 
\eqqn\dimlespa{
x = {\Delta m \over \Gamma} \, , \quad y = {\Delta \gamma \over 2\Gamma} \, ,
}
being $\Delta m = m_H - m_L$, $\Delta \gamma = \gamma_H - \gamma_L$ and 
$\Gamma = (\gamma_H + \gamma_L)/2$.
It is worth noting that the eight parameters
\eqqn\coeff{
\vcenter{\openup1\jot \halign{$\hfil#$&${}#\hfil$&
\qquad$\hfil#$&${}#\hfil$&
\qquad$\hfil#$&${}#\hfil$&
\qquad$\hfil#$&${}#\hfil$\cr
{\conj C}_y&=1 + \vert {\conj \xi}_F \vert^2,&
{\conj C}_x&=1 - \vert {\conj \xi}_F \vert^2,&
{\conj S}_y&=2{\rm Re} {\conj \xi}_F,&
{\conj S}_x&=-2{\rm Im} {\conj \xi}_F;\cr
{\conj {C^\prime}}_y&={\displaystyle{1 + \vert {\conj \xi}_F \vert^2 \over 
\vert \eta \vert^2}},&
{\conj {C^\prime}}_x&={\displaystyle-{1 - \vert {\conj \xi}_F \vert^2 \over 
\vert \eta \vert^2}},&
{\conj {S^\prime}}_y&={\displaystyle{2{\rm Re} {\conj \xi}_F \over 
\vertð\eta \vert^2}},&
{\conj {S^\prime}}_x&={\displaystyle{2{\rm Im} {\conj \xi}_F \over \vertð\eta \vert^2}};\cr}}
}
and the
coefficients of $f^{(\prime)}_i(t)$ and $h^{(\prime)}_i(t)$ 
in formulas \sevasym \ and
\sevasymII \ are all rephase--invariant quantities.
We apply the above general analyses to some specific cases. $CP$--violating 
asymmetries can be realized in semileptonic and nonleptonic decays.
Hadronic $CP$--asymmetries may be classified according to the final decay states.
Final states may be pure $CP$--eigenstates ($\ket{F}=CP\ket{F}=\ket{{\bar F}})$,
such as $F=\pi^+\pi^-, \pi^0\pi^0, \dots$ or $CP$--mixed states, such as
$K^0\pi^0, D^0\pi^0$ which can be recombined into $CP$--eigenstates. In the most
general case, both $F$ and ${\bar F}$ are common final states of $\bo$ and
$\bbar$, but they are not $CP$--eigenstates, such as $D^-\rho^+$.
A particular interest deserves the $CP$--violating asymmetry between
$\bo \rightarrow
F$ and  $\bbar \rightarrow {\overline F}$
(in the case  $F$ results a hadronic $CP$--eigenstate).
In this case the relevant asymmetry is
\eqqn\cpeig{
\eqalign{
a_{CP} = &{\Gamma ({\rm B}^0 (t) \rightarrow F) - \Gamma 
(\overline{\rm B}{}^0 (t) \rightarrow {\overline F}) \over 
\Gamma ({\rm B}^0 (t) \rightarrow F) + \Gamma 
(\overline{\rm B}{}^0 (t) \rightarrow {\overline F})} = 
{\left \vert f_{0{\overline 0}} {{\displaystyle 
A_{{\overline 0}F}}\over {\displaystyle A_{{\overline 0}{\overline F}}}} 
\right\vert^2 -1
\over {\left \vert f_{0{\overline 0}} {{\displaystyle 
A_{{\overline 0}F}}\over {\displaystyle A_{{\overline 0}{\overline F}}}} 
\right\vert^2 +1 }
}= \cr
%& \cr
= & {C^\Delta_y \cosh(y\tau) + C^\Delta_x \cos(x\tau) + 
S^\Delta_y \sinh(y\tau) + S^\Delta_x \sin(x\tau) \over 
C^\Sigma_y \cosh(y\tau) + C^\Sigma_x \cos(x\tau) + 
S^\Sigma_y \sinh(y\tau) + S^\Sigma_x \sin(x\tau)} \; .\cr}
}
The convenient notation $\tau = \Gamma t$ is used and the superscripts
$\Sigma$ and $\Delta$ denote, respectively, the sum or the difference between 
$C$ and $C^\prime$, $S$ and $S^\prime$.
The Standard Model predicts $\vert \eta \vert \simeq 1$ and, it holds to a good 
degree of accuracy. As a consequence, the following much simpler expression 
for Eq. \cpeig\ can be written as
\eqqn\approxmt{
a_{CP} \simeq
{[1 - \vert \xi_F\vert^2] \cos(x\tau) -2{\rm Im} \xi_F \sin(x\tau) 
\over
[1 + \vert \xi_F\vert^2] \cosh(y\tau) +2{\rm Re} \xi_F \sinh(y\tau)} \; .
}
Due to the smallness of $y$ ($y \simeq 0$), we derive the further standard 
result 
\eqqn\approx{
a_{CP} \simeq {\cal A} \cos( x \tau) - {\cal B} \sin (xð\tau) 
}
in terms of the two independent $CP$--violating parameters ${\cal A}$ and 
${\cal B}$ of Eq. \AeB. In the case of a 
nonvanishing ${\cal A}$, we speak of direct $CP$--violation in
the
decay amplitude. On the other side, the second parameter ${\cal B}$ vanishes
in the absence of $CP$--violating effects into $\bob$ mixing.
The importance of this expression is related 
to the possibility of performing a classification of the different 
forms of $CP$--violation in terms of ${\cal A}$ and 
${\cal B}$.
In the Standard Model, $CP$--violating effects can arise through the
interference between at least two independent amplitudes with different
$CP$--phases. Besides the charged W currents, 
some non zero $CP$--violating effects are provided by
loop induced transitions involving strong (gluon) and electroweak ($\gamma$,
$Z^0$ and $H^0$) interactions. 
Due to the presence of these penguin contributions, we cannot extract 
straightforwardly the $CP$--angle, which 
characterizes the $CP$--asymmetry between the two $CP$ conjugate decay modes.
In order to clarify this point, we decompose
the conjugated decay amplitudes $A_F$ and ${\overline A}_F$ as
\eqqn\amplitud{
\eqalign{
A_F = & A_1 e^{i \phi_1} e^{i \delta_1} + A_2 e^{i \phi_2} 
e^{i \delta_2} \cr
{\overline A}_F = & A_1 e^{-i \phi_1} e^{i \delta_1} + A_2 e^{-i \phi_2} 
e^{i \delta_2} \; , \cr}
}
where $\phi_1$, $\phi_2$ are weak phases associated with the different quark
mixing elements, whereas $\delta_1$, $\delta_2$ are unitarity
$CP$--conserving strong phases, usually, associated with the absorptive
part of the penguin diagrams or also related to inelastic final state
interactions. For convenience, we introduce the following 
quantities \eqqn\weakstron{ R = {A_2 \over A_1} \; , \quad \phi_{12} = \phi_1 -
\phi_2  \; , \quad \delta_{12} = \delta_1 - \delta_2 
}
being $A_1$ and $A_2$ the magnitudes of the hadronic 
matrix elements. Without any loss of generality, one may suppose
that a single weak
amplitude (or rather a single weak phase) dominates the decay process ($A_1 > 
A_2$). Therefore, 
the above parameters are found to be
\eqqn\abprmt{
\eqalign{
{\cal A} \simeq & -2R \sin \phi_{12} \sin \delta_{12} \cr
{\cal B} \simeq & -[\sin 2(\phi_M + \phi_1) - 2R\cos 2(\phi_M + \phi_1)
\sin \phi_{12} \cos \delta_{12}] \; , \cr}
}
where we have used the more
convenient notation
$\eta = - e^{-2i \phi_M}$ and
\eqqn\fnnf{
\eqalign{
\xi_F = & e^{-2i\phi_M} {{\overline A}_F \over A_F} = 
e^{-2i(\phi_M + \phi_1)}\left[ 1 + {2iR\sin \phi_{12} e^{-i \delta_{12}}
\over 1 + R e^{-i \phi_{12}} e^{-i \delta_{12}}} \right] \; , \cr
{\rm Im} \xi_F = & -\sin2(\phi_M + \phi_1) + \Delta_F  
\cr }
}
with $\Delta_F = -2R \sin\phi_{12} \cos \Big[ \delta_{12} + 2(\phi_M +\phi_1) 
\Big]$.
We first consider the particular case $R=0$ 
or $\phi_1 = \phi_2$. It is clear that it yields the box diagram dominant
result 
\eqqn\SMcontr{
\eqalign{
{\cal A} = & \, 0 \cr
{\cal B} \simeq & -\sin 2(\phi_M + \phi_1) \cr}
}
which is peculiar of indirect $CP$--violation in the mixing, and it is
due to the equality ${\overline A}_F = A^\ast_F$. 
The mixing-and-decay
$CP$--violating effects ($R\neq 0$, $\phi_1 \neq \phi_2$) induce a nonvanishing
${\cal A}$ (i.e. $\delta_{12} \neq 0$) with a direct $CP$--violation in the decay
amplitude  ($A_F$ or ${\overline A}_F$). As we have pointed out,
the existence of the direct $CP$--violation requires that both a strong and
a weak phase difference exist. 
Usually, we calculate the decay amplitudes by using the effective weak
Hamiltonian and the factorization approximation
\ref\MartBur{M. Ciuchini, E. Franco, G. Martinelli and L. Reina,
Nucl. Phys. {\bf B415} (1994) 403; G. Buchalla, A. J. Buras and 
M. E. Lautenbacher, Rev. Mod. Phys. {\bf 68} (1996) 1125}.
The evaluation of the relative strong phase $\delta_{12}$ is more
problematic as we lack a quantitative understanding of nonperturbative QCD
and of the effects due to the final state interactions or to the production of
coupled resonant decay modes.
Neglecting the final state interactions which produce the strong phase
difference $\delta_{12}$, we can consider the typical 
${\rm B}{}_d^0 \rightarrow \pi^+\pi^-$ decay process which is characterized 
by the
$CP$--angle $\alpha$ of the unitarity triangle \Pavia, being 
\eqqn\anng{
\sin 2\alpha = -\sin 2(\phi_M + \phi_1)
}
with
\eqqn\unttr{
\Delta_F = -R \left \{ { 2\cos \delta_{12} \sin \alpha + \sin 2\alpha 
[R - 2\cos (\alpha - \delta_{12})] \over 1 + R^2 - 2R \cos (\alpha - 
\delta_{12})} \right \}  \; .
}
Adopting the usual 
valence--quark convention, the ${\cal SU}(3)$ invariance can be used to 
isolate the gluon penguin contamination and determine $\alpha$ up to 
a two--fold ambiguity 
\ref\GroLon{
M. Gronau and D. London, Phys. Rev. Lett. {\bf 65} (1990) 3381}. 
The penguin effects on $CP$--violation are
supposed to be extracted by means of
the knowledge 
of relevant branching ratios, without observing the time--dependence of the decay 
rates.
Therefore, the $CP$--violation in the specimen case of
${\conj {\rm B}}{}_d^0 \rightarrow \pi^+\pi^-$ is
characterized by the following observables
\eqqn\speccas{
\eqalign{
{\cal A}_{\pi^+\pi^-} = & {1 - \vert \xi_{\pi^+\pi^-} \vert^2 \over 1 + \vert
\xi_{\pi^+\pi^-} \vert ^2} \cr
{\cal B}_{\pi^+\pi^-} = & {2 {\rm Im} \xi_{\pi^+\pi^-} \over 1 + \vert
\xi_{\pi^+\pi^-} \vert^2} \; , \cr}
}
where, in the isospin decomposition, we have that
\eqqn\isospin{
\eqalign{
\xi_{\pi^+\pi^-} = & e^{-2i\phi_M} \eta_{22} \left (
{1 +  \sqrt 2 \; {\displaystyle{ \eta_{02} \over \eta_{22}}}
\over 1 + \sqrt 2 \;
{\displaystyle{ \eta_{02} \over \eta_{00}}} } \right)= 
\cr
=&
\left({1 - {\widetilde \eta}_{22} \over 1 + {\widetilde \eta}_{22}}\right) \left(
{1 +  \sqrt 2 \;
{\displaystyle{{\widetilde \eta}_{02} \over {\widetilde \eta}_{00}}}
\; \left({\displaystyle {1 - {\widetilde \eta}_{00} \over 
1 - {\widetilde \eta}_{22}}}\right) \over 
1 + \sqrt 2 \;
{\displaystyle{{\widetilde \eta}_{02} \over {\widetilde \eta}_{00}}}
\; \left({\displaystyle {1 + {\widetilde \eta}_{00} \over 
1 + {\widetilde \eta}_{22}}}\right)}\right) \simeq 
%\cr \simeq & 
e^{2i\alpha} \left({1+{\overline z} \over 1+z }\right)
\cr}
}
with
\eqqn\states{
\eta_{I_1,I_2} = {A (\bbar \rightarrow \pi \pi, I=I_1) \over 
A(\bo \rightarrow \pi \pi, I=I_2)}  \; , \quad
{\widetilde \eta}_{I_1,I_2} = {A (B_H \rightarrow \pi \pi, I=I_1) \over 
A(B_L \rightarrow \pi \pi, I=I_2)} \; , 
}
being  $I_1,I_2 \in \{0,2\}$ the isospin states.
In this formula, we have introduced the quantities
\eqqn\zzbar{
{\conj z} = {\sqrt 2} \, 
{A \Big({\conj {\rm B}}{}^0 \rightarrow \pi \pi, I=0 \Big)
\over A \Big({\conj {\rm B}}{}^0 \rightarrow \pi \pi, I=2 \Big)} \;,
}
which are here represented by
\eqqn\zzz{
z = {\sqrt 2} \, {\eta_{02} \over \eta_{00}} \qquad {\rm and} \qquad
{\overline z} = {\sqrt 2} \, {\eta_{02} \over \eta_{22}} \; .
}
For such a decay mode, the relevant $CP$--violating observable becomes
${\rm Im} \xi_{\pi^+\pi^-} =  \sin 2 \alpha + \Delta_{+-}$,
where $\alpha$ in terms of the Wolfenstein $\rho - \eta$ parameters \Pavia\ is
given by
\eqqn\Wolfen{
\alpha = {\rm arg} \left( - {V_{td} V^{\ast}_{tb} \over V_{ud}
V^{\ast}_{ub}} \right) = 
\arctan \left( {\eta \over \rho (\rho - 1) + \eta^2}
\right) \; , 
}
and
\eqqn\VIbis{
\Delta_{+-}
\simeq -2R \biggl[\cos \delta_{12} \sin \alpha - 
\sin 2\alpha \cos (\alpha - \delta_{12}) \biggr]  \; .
}
However, experimentally, it may be difficult to measure the deviation 
$\Delta_{+-}$ accurately, because it requires the measurement of the 
difficult decay ${\rm B}{}_d^0 \rightarrow \pi^0 \pi^0$, whose branching ratio
is expected to be very small ($\simeq 10^{-6}$), due to color suppression
\ref\Kramer{
G. Kramer and W. F. Palmer, Phys. Rev. {\bf D52} (1995) 6411}.
Within the flavour and isospin invariance, it has been pointed out that the
isospin relations for amplitude differences in ${\rm B} \rightarrow \pi {\rm
K}$  
\ref\DESIII{
M. Gronau, J. L. Rosner and D. London , Phys. Rev. Lett. {\bf 73} (1994) 21;
J. P. Silva and L. Wolfenstein, Phys. Rev. {\bf D49} (1994) 1151;
N. G. Deshpande and X.-G. He, Phys. Rev. Lett. {\bf 75} (1995) 1703}\
may improve the situation. Furthermore, the inclusion of the 
${\rm B} \rightarrow \pi \rho$ and other decay modes, can also remove the
gluon penguin contamination and solve the two--fold ambiguity
\ref\Gross{
R. Aleksan \etal, Phys. Lett. {\bf B356} (1995) 95;
Y. Grossman and H. R. Quinn, Phys. Rev. {\bf D 56} (1997) 7259 and refs.
therein},
by studying the full Dalitz plot and the time--dependence for $3 \pi$ decay
channels
\ref\ADKLD{
R. Aleksan, I. Dunietz, B. Kayser and F. Le Diberdier,
Nucl. Phys. {\bf B361} (1991) 141}.
The reliability of all these approaches is limited by possible ${\cal SU}$(3)
breaking effects originating, for instance, from the presence of electroweak
penguins which become important due to the fact that the  $Z^0$ exchange
depends on the square of the top quark mass
\ref\Fleish{R. Fleisher, Int. J. Mod. Phys. {\bf A12} (1997) 2459}.
Indeed, without any deep insight into the world of the penguin operators
and into other isospin breaking effects 
\ref\Deshp{N. G. Deshpande and X. G. He, Phys. Rev. Lett. {\bf 74} (1995)
26; 
M. Gronau, O. F. Hern\'andez, D. London and J. L. Rosner, Phys.
Rev. {\bf D52} (1995) 6374;
R. Fleisher, Phys. Lett. {\bf B365} (1996) 399;
N. G. Deshpande, X. G. He and S. Oh, Phys. Lett. {\bf B384} (1996) 283;
R. Fleischer and T. Mannel, Phys. Lett. {\bf B397} (1997) 269;
N. G. Deshpande, X. G. He and S. Oh, Z. Phys. {\bf C74} (1997) 359
},
we can simply state that the influence of ${\cal SU}$(3)--breaking effects 
(especially electroweak penguins) on the extraction of $\alpha$
is rather controversial \Deshp.
The deviation $\Delta_{+-}$ in Eq. \VIbis\ seems sizeable as large as $0.12$
using only factorization approximation. However, it is expected to be much
smaller, if we include the further constraint
$\vert A({\rm B}^- \rightarrow \pi^- \pi^0 ) \vert =  
\vert A({\rm B}^+ \rightarrow \pi^+ \pi^0 ) \vert$
coming from the study of the decay rates of the charged
${\rm B}$ mesons \Deshp.

\noindent
A better understanding of the $CP$--violating effects in $B$ meson
decays originate from a full--fledged knowledge of the interference
of the tree level amplitude with higher
order corrections to vertex and masses. 
Concluding the paper, however, it is worth noting that this picture is obscured
by other intriguing effects which support our need to consider a 
model--independent analysis. 
We mean the problems of the relative strong phase $\delta_{12}$
and the peak-dip structure which emerge when we consider the exchange of two
(or more) resonances. The latter resembles a $s$-channel interaction between
the initial and final states, and can be described by means of a
$q^2$--dependent width in the Breit-Wigner intermediate propagators
\ref\CPWIDTH{ A. Pilaftsis and M. Nowakowski, Int. J. Mod. Phys. {\bf A9}
(1994) 1097, (E) 5849; G. Eilam, M. Gronau and R. Mendel, Phys. Rev. Lett.
{\bf 74} (1995) 4984}. The particle width effects of the intermediate propagators
should be taken into account since they introduce a $CP$ odd contribution. In
this context, it is also possible to consider the interference effects
in the breaking of some relevant isospin relations induced by the intermediate
$\rho$--$\omega$ mixing
\ref\ROMIX{R. Enomoto and M. Tanabashi, Phys. Lett. {\bf B386} (1996) 413;
S. Gardner, H. B. O'Connell, A. W. Thomas , Phys. Rev. Lett. {\bf 80}
(1998) 1834}.
On the other side,
the question of how to calculate the relative strong phase $\delta_{12}$
has been discussed both at the quark and at hadronic level. At the quark
level, the necessary strong phases depend on the absorptive parts of
$t$-channel vertex corrections (HARD FSI). These contributions are provided by
different loop effects based both on the loop quark--rescattering (time like
penguins) \ref\TPEN{J. M. Gerard and W. S. Hou, Phys. Rev. Lett. {\bf 62}
(1989) 855; Phys. Rev. {\bf D43} (1991) 2909; H. Simma, G. Eilam and D. Wyler,
Nucl. Phys. {\bf B352} (1991) 367}, and on final state hadronization
(space-like penguins) 
{\ref\SPEN{M. Tanimoto \etal\ Phys. Rev. {\bf D42} (1990) 252;
D.-S. Du, M.-Z. Yang, D.-Z. Zhang, Phys. Rev. {\bf D53} (1996) 249},
in dependence of the
gluonic momentum transfer. This quark level approach is related to the
influence of virtual gluons in the form of the interactions, and therefore,
it is intimately connected to CPT--invariance.
In fact, the stringent CPT
constraints can be evaded by a partial sum of final states. Also the
factorization consistency \ref\Facto{N. Cabibbo and L. Maiani, Phys. Lett. {\bf
B73} (1978) 418; J. D. Bjorken, Nucl. Phys. (Proc. Suppl.) {\bf 11} (1989) 287}\
is complicated by the presence of the strong rescattering of the intermediate
virtual states (with the same quark gluon content of the final decay modes)
just because they violate the cancellations imposed by unitarity and
CPT--invariances
\ref\soares{
J. M. Soares, Phys. Rev. Lett. {\bf 79} (1997) 1166}.
Therefore, the danger of a nonvanishing FSI cannot be avoided if
the factorization ansatz cannot be done. At hadronic level, the plot thickens
when we consider the inelastic $t$-channel interactions (SOFT FSI) which
induce the mixing of final decay products. Usually, final coupled decay
modes are considered by means of a parametrization of the strong
$S$--rescattering matrix with the pomeron dominance and Regge model.
The S-matrix of two coupled resonant channels in B-meson decays is probably
negligible
\ref\RESMAT{M. Wanninger and L. M. Sehgal, Z. Phys. {\bf C50} (1991) 47;
B. Blok, M. Gronau and J. Rosner, Phys. Rev. Lett. {\bf 78} (1997) 3999},
but in general the phase shift effects depend strongly on the kinematical
configuration of the particles
\ref\WYL{M. Simonius and D. Wyler, Z. Phys. {\bf C42} (1989) 471;
C. Greub, H. Simma and D. Wyler, Nucl. Phys. {\bf B434} (1995) 39}.
Finally, we can say that
although, $CP$--violation has been observed only in
the $\kok$ complex  system, large $CP$--violating effects are expected in the
${ {\rm B}_d^ 0 - \overline{\rm B}{}_d^0 }$ system. 
As we mentioned, the charge asymmetry in semileptonic decays Eq. \esl\ 
is predicted to be very small, without the inclusion of
new physical effects in $\bob$ mixing 
\cd . But, in the nonleptonic 
decays, the relative asymmetries may be large due 
to the interplay of mixing and decay amplitude.
This problem was widely discussed in the 
case the hadronic final states $F$ being $CP$--eigenstates.
The predictions for the partial decay rate asymmetries seem to put in
evidence some $CP$ breaking effects which, presumably,
are sensitive to the strong interactions.
In this letter, we propose a scattering theory to describe the $CP$--violation
and we provide a complete set of the rephasing invariant observables. 
In particular, our proposal becomes a valid tool to provide the characterization
of the $CP$--violating effects in neutral $B$ decays in two pseudoscalar
mesons,
and, in the same time, it can parameterize the difficult problem of the
penguin unreliabilities.

\vfil\eject
\vfill\eject\immediate\closeout\rfile%\parindent=20pt
\centerline{{\bf References}}\bigskip% \frenchspacing%
\input refs.tmp\vfill\eject
 \bye